\documentclass[twocolumn,pre]{revtex4}

\usepackage{dcolumn}
\usepackage{amsmath}

\usepackage{graphicx}
\usepackage{rotating}
\usepackage{multirow}

\begin{document}

\renewcommand{\d}{{\rm d}}
\renewcommand{\i}{{\rm i}}
\renewcommand{\O}{{\rm O}}
\newcommand{\e}{{\rm e}}
\newcommand{\defn}{\textit}
\newcommand{\half}{\mbox{$\frac12$}}
\newcommand{\set}[1]{\lbrace#1\rbrace}
\newcommand{\av}[1]{\langle#1\rangle}
\newcommand{\eref}[1]{(\ref{#1})}
\newcommand{\etal}{{\it{}et~al.}}
\newcommand{\cF}{\mathcal{F}}
\newcommand{\cG}{\mathcal{G}}
\newcommand{\Tr}{\mathop{\rm Tr}}
\newcommand{\cov}{\mathop{\rm cov}}
\newcommand{\var}{\mathop{\rm var}}
\newcommand{\Li}{\mathop{\rm Li}}
\newcommand{\norm}[1]{\|\,#1\,\|}
\newcommand{\phm}{\phantom{-}}

\newlength{\figurewidth}
\ifdim\columnwidth<10.5cm
  \setlength{\figurewidth}{0.95\columnwidth}
\else
  \setlength{\figurewidth}{10cm}
\fi
\setlength{\parskip}{0pt}
\setlength{\tabcolsep}{6pt}
\setlength{\arraycolsep}{2pt}

\title{Properties of highly clustered networks}
\author{M. E. J. Newman}
\affiliation{Department of Physics and Center for the Study of Complex
Systems,\\
University of Michigan, Ann Arbor, MI 48109--1120}
\begin{abstract}
We propose and solve exactly a model of a network that has both a tunable
degree distribution and a tunable clustering coefficient.  Among other
things, our results indicate that increased clustering leads to a decrease
in the size of the giant component of the network.  We also study SIR-type
epidemic processes within the model and find that clustering decreases the
size of epidemics, but also decreases the epidemic threshold, making it
easier for diseases to spread.  In addition, clustering causes epidemics to
saturate sooner, meaning that they infect a near-maximal fraction of the
network for quite low transmission rates.
\end{abstract}
\maketitle

\section{Introduction}
There has in recent years been considerable interest within the physics
community in the structure and dynamics of networks, with applications to
the Internet, the World-Wide Web, citation networks, and social and
biological networks~\cite{Strogatz01,AB02,DM02}.  Two significant
properties of networks have been particularly highlighted.  First, one
observes for most networks that the degree distribution is highly
non-Poissonian~\cite{AJB99,FFF99,ASBS00,Newman01a,Liljeros01}.  (A network
consists of a set of nodes or ``vertices'', joined by lines or ``edges'',
and the degree of a vertex is the number of edges attached to that vertex.)
Histograms of vertex degree for many networks show a power-law form with
exponent typically between $-2$ and~$-3$, while other networks may have
exponential or truncated power-law distributions.  Second, it is found that
most networks have a high degree of transitivity or clustering, i.e.,~that
there is a high probability that ``the friend of my friend is also my
friend''~\cite{WS98}.  In topological terms, this means that there is a
heightened density of loops of length three in the network, and more
generally it is found that networks have a heightened density of short
loops of various lengths~\cite{CPV03}.

It is now well understood how to calculate the properties of networks with
arbitrary degree distributions~\cite{MR95,MR98,ACL00,NSW01,CL02a}, but
where clustering is concerned our understanding is much poorer.  Most of
the standard techniques used to solve network models break down when
clustering is introduced, obliging researchers to turn to numerical
methods~\cite{WS98,KE02,HK02b,MV03}.

In this paper, we present a plausible network model that incorporates both
non-Poisson degree distributions and non-trivial clustering, and which is
exactly solvable for many of its properties, including component sizes,
percolation threshold, and clustering coefficient.  Our results show that
clustering can have a substantial effect on the large-scale structure of
networks, and produces behaviors that are both quantitatively and
qualitatively different from the simple non-clustered case.

The outline of the paper is as follows.  In Sec.~\ref{secmodel} we define
our model and in Sec.~\ref{analytics} we derive exact expressions for a
variety of its properties.  In Sec.~\ref{results} we discuss the form of
these expressions for some sensible choices of the parameters, and also
consider the behavior of epidemic processes within our model.  In
Sec.~\ref{concs} we give our conclusions.

\section{The model}
\label{secmodel}
There is empirical evidence that clustering in networks arises because the
vertices are divided into groups~\cite{DGM02a,RB03}, with a high density of
edges between members of the same group, and hence a high density of
triangles, even though the density of edges in the network as a whole may
be low.  Our model is perhaps the simplest and most obvious realization of
this idea.  We describe it here in the anthropomorphic language of social
networks, although our arguments apply equally to non-social networks.

We consider a network of $N$ individuals divided into $M$ groups.  A social
network, for example, might be divided up according to the location,
interests, occupation, and so forth of its members.  (Many networks are
indeed known to be divided into such groups~\cite{GN02}.)  Individuals can
belong to more than one group, the groups they belong to being chosen---in
our model---at random.  Individuals are not necessarily acquainted with all
other members of their groups.  If two individuals belong to the same group
then there is a probability~$p$ that they are acquainted and $q=1-p$ that
they are not; if they have no groups in common then they are not
acquainted.  (A more sophisticated model in which there are many nested
levels of groups within groups and a spectrum of acquaintance probabilities
depending on these levels has been proposed and studied numerically by
Watts~\etal~\cite{WDN02}.  For this paper, however, we confine ourselves to
the simpler case.)  In addition to the probability~$p$, the model is
parametrized by two probability distributions: $r_m$~is the probability
that an individual belongs to $m$ groups and $s_n$ is the probability that
a group contains $n$ individuals.

\begin{figure}
\begin{center}
\resizebox{\figurewidth}{!}{\includegraphics{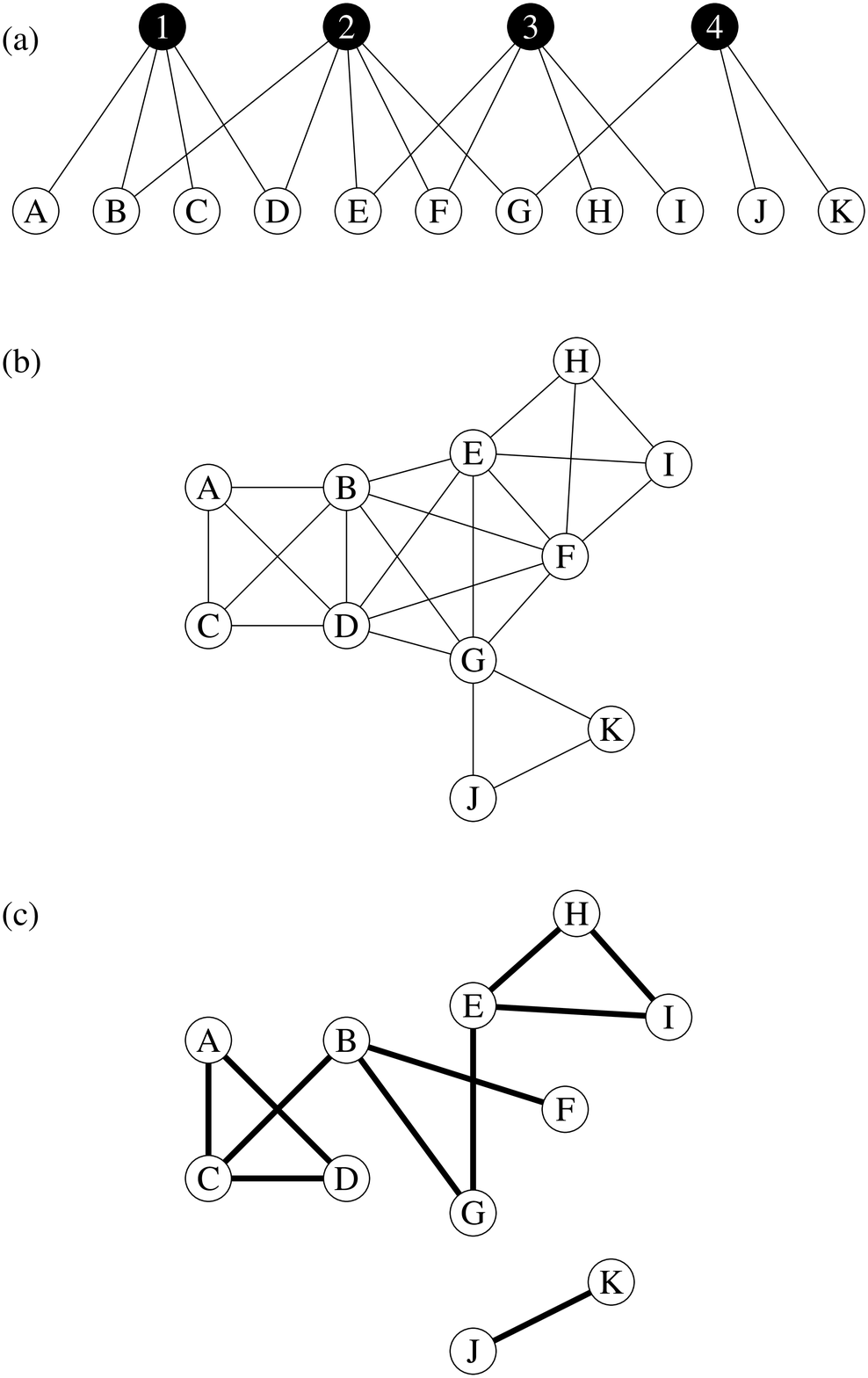}}
\end{center}
\caption{The structure of the network model described in this paper.
(a)~We represent individuals (A--K) and the groups (1--4) to which they
belong with a bipartite graph structure.  (b)~The bipartite graph is
projected onto the individuals only.  (c)~The connections between
individuals are chosen by bond percolation on this projection with bond
occupation probability~$p$.  The net result is that individuals have
probability $p$ of knowing others with whom they share a group.}
\label{model}
\end{figure}

Mathematically, the model can be regarded as a bond percolation model on
the one-mode projection of a bipartite random graph.  The structure of
individuals and groups forms the bipartite graph, the network of shared
groups is the projection of that graph onto the individuals alone, and the
probability~$p$ that one of the possible contacts in this projection is
actually realized corresponds to a bond percolation process on the
projection.  See Fig.~\ref{model}.

\section{Analytic developments}
\label{analytics}
We can derive a variety of exact results for our model in the limit of
large size using generating function methods.  There are four fundamental
generating functions that we will use:
\begin{eqnarray}
\label{defsf0f1}
\!\! f_0(z) &=& \! \sum_{m=0}^\infty r_m z^m,\hspace{8.0pt}
     f_1(z)  =  \mu^{-1} \sum_{m=0}^\infty m r_m z^{m-1},\\
\label{defsg0g1}
\!\! g_0(z) &=& \sum_{n=0}^\infty s_n z^n,\hspace{12.2pt}
     g_1(z)  =  \nu^{-1} \sum_{n=0}^\infty n s_n z^{n-1},
\end{eqnarray}
where $\mu=\sum_m m r_m$ and $\nu=\sum_n n s_n$ are the mean numbers of
groups per person and people per group respectively.

\subsection{Degree distribution}
Consider a randomly chosen person~A, who belongs to some number of
groups~$m$.  The number~$j$ of A's acquaintances within one particular
group of size~$n$ is binomially distributed according to ${n-1\choose j}
p^j q^{n-1-j}$.  We represent this distribution by its generating function:
\begin{equation}
\sum_{j=0}^{n-1} {n-1\choose j} p^j q^{n-1-j} z^j = [pz+q]^{n-1}.
\end{equation}
Averaging over group size, the full generating function for neighbors in a
single group is $\nu^{-1}\sum_{n=0}^\infty n s_n [pz+q]^{n-1}=g_1(pz+q)$,
and for neighbors of a single person is $f_0(g_1(pz+q))$.  This allows us
to calculate the degree distribution for any given $\set{r_m,s_n}$, and by
judicious choice of the fundamental distributions, we can arrange for the
degree distribution to take a wide variety of forms.  We give some examples
shortly.  The mean degree~$\av{k}$ of an individual in the network is given
by
\begin{equation}
\av{k} = \bigl[ {\partial_z} f_0(g_1(pz+q)) \bigr]_{z=1}\!\!
       = p\mu g_1'(1).
\label{avk}
\end{equation}

\begin{table*}[t]
\begin{tabular}{l|l}
$k$ & $P(k|k)$ \\
\hline
1 & $1$ \\

2 & $p$ \\

3 & $3\,p^2 q + p^3$ \\

4 & $16\,p^3 q^3 + 15\,p^4 q^2 + 6\,p^5 q + p^6$ \\

5 & $125\,p^4 q^6 + 222\,p^5 q^5 + 205\,p^6 q^4 + 120\,p^7 q^3
      + 45\,p^8 q^2 + 10\,p^9 q + p^{10}$ \\

6 & $1296\,p^5 q^{10} + 3660\,p^6 q^9 + 5700\,p^7 q^8 + 6165\,p^8 q^7
      + 4945\,p^9 q^6 + 2997\,p^{10} q^5 + 1365\,p^{11} q^4
      + 455\,p^{12} q^3 + 105\,p^{13} q^2$ \\
  & \qquad ${} + 15\,p^{14} q + p^{15}$ \\

7 & $16807\,p^6 q^{15} + 68295\,p^7 q^{14} + 156555\,p^8 q^{13}
      + 258125\,p^9 q^{12} + 331506\,p^{10} q^{11}
      + 343140\,p^{11} q^{10} + 290745\,p^{12} q^9$ \\
  & \qquad ${} + 202755\,p^{13} q^8 + 116175\,p^{14} q^7 + 54257\,p^{15} q^6
      + 20349\,p^{16} q^5 + 5985\,p^{17} q^4 + 1330\,p^{18} q^3
      + 210\,p^{19} q^2 + 21\,p^{20} q + p^{21}$ \\

8 & $262144\,p^7 q^{21} + 1436568\,p^8 q^{20} + 4483360\,p^9 q^{19} +
    10230360\,p^{10} q^{18} + 18602136\,p^{11} q^{17} + 28044072\,p^{12}
    q^{16}$ \\
  & \qquad ${} + 35804384\,p^{13} q^{15} + 39183840\,p^{14} q^{14} +
    37007656\,p^{15} q^{13} + 30258935\,p^{16} q^{12} + 21426300\,p^{17}
    q^{11} + 13112470\,p^{18} q^{10}$ \\
  & \qquad ${} + 6905220\,p^{19} q^9 + 3107937\,p^{20} q^8 +
1184032\,p^{21} q^7 + 376740\,p^{22} q^6 + 98280\,p^{23} q^5 +
20475\,p^{24} q^4 + 3276\,p^{25} q^3$ \\
  & \qquad ${} + 378\,p^{26} q^2 + 28\,p^{27} q + p^{28}$ \\

9 & $4782969\,p^8 q^{28} + 33779340\,p^9 q^{27} + 136368414\,p^{10} q^{26}
    + 405918324\,p^{11} q^{25} + 974679363\,p^{12} q^{24} +
    1969994376\,p^{13} q^{23}$ \\
  & \qquad ${} + 3431889000\,p^{14} q^{22} + 5228627544\,p^{15} q^{21} +
    7032842901\,p^{16} q^{20} + 8403710364\,p^{17} q^{19} +
    8956859646\,p^{18} q^{18}$ \\
  & \qquad ${} + 8535294180\,p^{19} q^{17} + 7279892361\,p^{20} q^{16} +
    5557245480\,p^{21} q^{15} + 3792906504\,p^{22} q^{14} +
    2309905080\,p^{23} q^{13}$ \\
  & \qquad ${} + 1251493425\,p^{24} q^{12} + 600775812\,p^{25} q^{11} +
    254183454\,p^{26} q^{10} + 94143028\,p^{27} q^9 + 30260331\,p^{28} q^8 +
    8347680\,p^{29} q^7$ \\
  & \qquad ${} + 1947792\,p^{30} q^6 + 376992\,p^{31} q^5 +
58905\,p^{32} q^4 + 7140\,p^{33} q^3 + 630\,p^{34} q^2 + 36\,p^{35} q +
p^{36}$ \\

10 & $100000000\,p^9 q^{36} + 880107840\,p^{10} q^{35} + 4432075200\,p^{11}
     q^{34} + 16530124800\,p^{12} q^{33} + 50088981600\,p^{13} q^{32}$ \\
   & \qquad ${} + 128916045720\,p^{14} q^{31} + 288982989000\,p^{15} q^{30} +
     573177986865\,p^{16} q^{29} + 1016662746825\,p^{17} q^{28}$ \\
   & \qquad ${} + 1624745199910\,p^{18} q^{27} + 2352103292070\,p^{19} q^{26} +
     3096620034795\,p^{20} q^{25} + 3717889913655\,p^{21} q^{24}$ \\
   & \qquad ${} + 4078716030900\,p^{22} q^{23} + 4093594934220\,p^{23} q^{22} +
     3761135471805\,p^{24} q^{21} + 3163862003211\,p^{25} q^{20}$ \\
   & \qquad ${} + 2435820178050\,p^{26} q^{19} + 1714943046390\,p^{27} q^{18} +
     1102765999275\,p^{28} q^{17} + 646542946125\,p^{29} q^{16}$ \\
   & \qquad ${} + 344847947664\,p^{30} q^{15} + 166867565040\,p^{31} q^{14} +
     73005619995\,p^{32} q^{13} + 28759950345\,p^{33} q^{12} +
     10150589610\,p^{34} q^{11}$ \\
   & \qquad ${} + 3190186926\,p^{35} q^{10} + 886163125\,p^{36}
     q^9 + 215553195\,p^{37} q^8 + 45379620\,p^{38} q^7 + 8145060\,p^{39} q^6 +
     1221759\,p^{40} q^5$ \\
   & \qquad ${} + 148995\,p^{41} q^4 + 14190\,p^{42} q^3 + 990\,p^{43}
     q^2 + 45\,p^{44} q + p^{45}$

\end{tabular}
\caption{The polynomials $P(k|k)$ for values of $k$ up to~10.}
\label{tablepll}
\end{table*}

\subsection{Clustering coefficient}
The clustering coefficient~$C$ is a measure of the level of clustering in a
network~\cite{WS98}.  It is defined as the mean probability that two
vertices in a network are connected, given that they share a common network
neighbor.  Mathematically it can be written as three times the ratio of the
number of triangles~$N_\triangle$ in the network to the number of connected
triples of vertices~$N_3$~\cite{NSW01}.  In the present case, we have
\begin{eqnarray}
N_\triangle &=& \mbox{$\frac16$} N p^3 f_0'(1) g_1''(1),\nonumber\\
N_3         &=& \half N p^2 \bigl[ f_0''(1) [g_1'(1)]^2 + f_0'(1) g_1''(1)
                \bigr],
\end{eqnarray}
and hence the clustering coefficient is
\begin{equation}
C = {3N_\triangle\over N_3} = p {f_0'(1) g_1''(1)\over
                         f_0''(1) [g_1'(1)]^2 + f_0'(1) g_1''(1)}
  = pC_b,
\label{cvalue}
\end{equation}
where $C_b$ is the clustering coefficient of the simple one-mode projection
of the bipartite graph, Fig.~\ref{model}b~\cite{NSW01}.  In other words,
one can interpolate smoothly and linearly from $C=0$ to the maximum
possible value for this type of graph, simply by varying~$p$.  (In the
limit $C=0$ our model becomes equivalent to the standard unclustered random
graphs studied previously~\cite{MR95,NSW01}.)  The average number of groups
to which people belong and the parameter~$p$ give us two independent
parameters that we can vary to allow us to change $C$ while keeping the
mean degree~$\av{k}$ constant.  Alternatively, and perhaps more logically,
we can regard $C$ and $\av{k}$ as the defining parameters for the model and
calculate the appropriate values of other quantities from these.

The \emph{local} clustering coefficient~$C_i$ for a vertex~$i$ has also
been the subject of recent study.  $C_i$~is defined to be the fraction of
pairs of neighbors of~$i$ that are neighbors also of each
other~\cite{WS98}.  For a variety of real-world networks $C_i$ is found to
fall off with the degree~$k_i$ of the vertex as $C_i\sim
k_i^{-1}$~\cite{DGM02a,RB03}.  This behavior is reproduced nicely by our
model.  Vertices with higher degree belong to more groups in proportion to
$k_i$ while the number of pairs of their neighbors is $\half k_i(k_i-1)$,
and the combination gives precisely $C_i\sim k_i^{-1}$ as $k_i$ becomes
large.

\subsection{Component structure}
To solve for the component structure of the model we focus on acquaintance
patterns within a single group.  Suppose person~A belongs to a group of $n$
people.  We would like to know how many individuals within that group A is
connected to, either directly (via a single edge) or indirectly (via any
path through other members of the group).  Let $P(k|n)$ be the probability
that vertex A belongs to a connected cluster of $k$ vertices in the group,
including itself.  We have
\begin{equation}
P(k|n) = {n-1\choose k-1} q^{k(n-k)} P(k|k),
\label{defsplk}
\end{equation}
which follows since we can make an appropriate graph of $n$ labeled
vertices by taking a graph of $k$ vertices, to all of which A is connected,
and adding $n-k$ others to it, which we can do in ${n-1\choose k-1}$
distinct ways, each with probability $q^{k(n-k)}$ (the probability that
none of the newly added vertices connects to any of the $k$ old vertices).

The probabilities $P(k|k)$ are polynomials in $p$ of order $s=\half k(k-1)$
that can be written in the form
\begin{equation}
P(k|k) = \sum_{l=0}^s M^k_l p^l q^{s-l},
\label{pkform}
\end{equation}
where $M^k_l$ is the number of labeled connected graphs with $k$ vertices
and $l$ edges.  While some progress can be made in evaluating the $M^k_l$
by analytic methods (see Appendix~A), the resulting expressions are poorly
suited to mechanical enumeration of~$P(k|k)$.  For practical purposes, it
is simpler to observe that
\begin{equation}
P(k|k) = 1 - \sum_{l=0}^{k-1} P(l|k),
\label{normalize}
\end{equation}
which in combination with Eq.~\eref{defsplk} allows us to evaluate $P(k|k)$
iteratively, given the initial condition $P(1|1)=1$.  In
Table~\ref{tablepll} we give the first few $P(k|k)$ for $k$ up to~10.

The generating function for the number of vertices to which A is connected,
by virtue of belonging to this group of size~$n$, is:
\begin{eqnarray}
h_n(z) &=& \sum_{k=1}^n P(k|n) \,z^{k-1}\nonumber\\
       &=& \sum_{k=1}^n {n-1\choose k-1} q^{k(n-k)} P(k|k) \,z^{k-1}.
\label{defshn}
\end{eqnarray}
Notice the appearance of $z^{k-1}$---this is a generating function for the
number of vertices A is connected to excluding itself.  Averaging over the
size distribution of groups then gives $h(z) = \nu^{-1}
\sum_n n s_n h_n(z)$, and the total number of others to whom A is connected
via all the groups they belong to is generated by $G_0(z)=f_0(h(z))$, where
$f_0(z)$ is defined in Eq.~\eref{defsf0f1}.  If we reach an individual by
following a randomly chosen edge, then we are more likely to arrive at
individuals who belong to a large number of groups.  This means that the
distribution of other groups to which such an individual belongs is
generated by the function $f_1(z)$ in Eq.~\eref{defsf0f1}, and the number
of other individuals to which they are connected is generated by
$G_1(z)=f_1(h(z))$.

Armed with these results, we can now calculate a variety of quantities for
our model.  We focus on two in particular, the position of the percolation
threshold and the size of the giant component.  The distribution of the
number of individuals one step away from person~A is generated by the
function~$G_0(z)$, while the number two steps away is generated by
$G_0(G_1(z))$.  There is a giant component in the network if and only if
the average number two steps away exceeds the average number one step
away~\cite{NSW01}.  (This is a natural criterion: it implies that the
number of people reachable is increasing with distance.)  Thus there is a
giant component if $\bigl[ \partial_z \bigl( G_0(G_1(z)) - G_0(z) \bigr)
\bigr]_{z=1} > 0$.  Substituting for $G_0$ and~$G_1$, this result can be
written
\begin{equation}
f_1'(1)\,h'(1) > 1.
\label{perccond}
\end{equation}

When this condition is satisfied and there is a giant component, we define
$u$ to be the probability that one of the individuals to whom A is
connected is \emph{not} a member of this giant component.  A~is also not a
member provided all of its neighbors are not, so that $u$ satisfies the
self-consistency condition $u=G_1(u)$.  Then the size of the giant
component is given by $S=1-G_0(u)$.

\section{Results}
\label{results}
As an example of the application of these results, consider the simple
version of our model in which all groups have the same size~$n=\nu$.  Then
$h(z)=h_\nu(z)$ and the degree distribution is dictated solely by the
distribution~$r_m$ of the number of groups to which individuals belong.  We
consider two examples of this distribution, a Poisson distribution and a
power-law distribution.

Let us look first at the Poisson case $r_m=\mu^m \e^{-\mu}/m!$, for which
the calculations are particularly simple.  The Poisson distribution
corresponds to choosing the members of each group independently and
uniformly at random.  From Eqs.~\eref{avk} and~\eref{cvalue} we have
\begin{equation}
\av{k} = p\mu(\nu-1),\quad
     C = {p\over 1 + \mu(\nu-1)/(\nu-2)}.
\end{equation}

\begin{figure}
\begin{center}
\resizebox{\figurewidth}{!}{\includegraphics{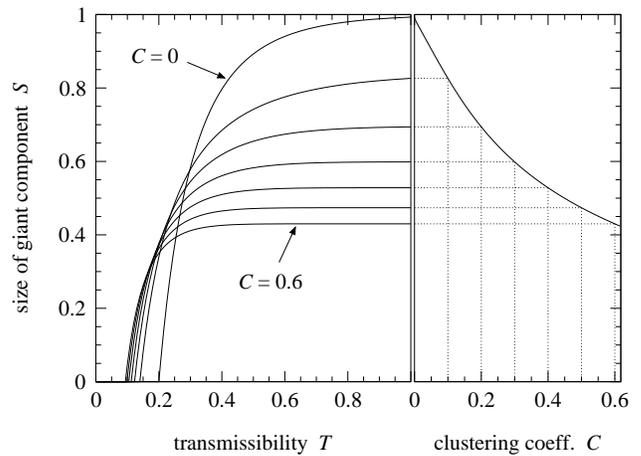}}
\end{center}
\caption{Right panel: the size of the giant component of the graph as a
function of clustering coefficient for the Poisson case with group size
$\nu=10$ and mean degree $\av{k}=5$.  Left panel: the size of an epidemic
outbreak for an SIR model on our network as a function of
transmissibility~$T$, for values of $C$ from $0$ to $0.6$ in steps
of~$0.1$.}
\label{poisson}
\end{figure}

In the right-hand panel of Fig.~\ref{poisson} we show results for the size
of the giant component as a function of clustering for the case of groups
of size $\nu=10$ with $\av{k}=5$.  As the figure shows, the giant component
size decreases sharply as clustering is increased.  The physical insight
behind this result is that for given~$\av{k}$, high clustering means that
there are more edges in all components, including the giant component, than
are strictly necessary to hold the component together---there are many
redundant paths between vertices formed by the many short loops of edges.
Since fixing~$\av{k}$ also fixes the total number of edges, this means that
the components must get smaller; the redundant edges are in a sense wasted,
and the percolation properties of the network are similar to those for a
network with fewer edges.

\subsection{Epidemics}
A topic of particular interest in the recent literature has been the spread
of disease over networks.  The classic SIR model of epidemic
disease~\cite{Hethcote00} can be generalized to an arbitrary contact
network, and maps onto a bond percolation model on that network with bond
occupation probability equal to the transmissibility~$T$ of the
disease~\cite{Grassberger83,Sander02}.  Since we have already solved the
bond percolation problem for our networks, we can also immediately solve
the SIR model, by making the substitution $p\to pT$.  We show some results
in the left-hand panel of Fig.~\ref{poisson} for the same choice of degree
distributions as before.  In general we see a percolation transition at
some value of~$T$, which corresponds to the epidemic threshold for the
model (denoted $R_0=1$ in traditional mathematical epidemiology).  Above
this threshold there is a giant component whose size measures the number of
people infected in an epidemic outbreak of the disease.

The size of the epidemic tends to the size of the giant component for the
network as a whole as $T\to1$, as represented by the dotted lines in the
figure, and is therefore typically smaller the higher the value of the
clustering coefficient.  However, it is interesting to note also that
as~$C$ becomes large the epidemic size saturates long before~$T=1$,
suggesting that in clustered networks epidemics will reach most of the
people who are reachable even for transmissibilities that are only slightly
above the epidemic threshold.  This behavior stands in sharp contrast to
the behavior of ordinary fully mixed epidemic models, or models on random
graphs without clustering, for which epidemic size shows no such
saturation~\cite{Hethcote00,Newman02c}.  It arises precisely because of the
many redundant paths between individuals introduced by the clustering in
the network, which provide many routes for transmission of the disease,
making it likely that most individuals who can catch the disease will
encounter it by one route or another, even for quite moderate values
of~$T$.

As we can also see from Fig.~\ref{poisson}, the position of the epidemic
threshold \emph{decreases} with increasing clustering.  At first this
result appears counter-intuitive.  The smaller giant component for higher
values of~$C$ seems to indicate that the model finds it harder to
percolate, and we might therefore expect the percolation threshold to be
higher.  In fact, however, the many redundant paths between vertices when
clustering is high make it easier for the disease to spread, not harder,
and so lower the position of the threshold.  Thus clustering has both bad
and good sides were the spread of disease is concerned.  On the one hand
clustering lowers the epidemic threshold for a disease and also allows the
disease to saturate the population at quite low values of the
transmissibility, but on the other hand the total number of people infected
is decreased.

\subsection{Power-law degree distributions}
Now consider the case of a power-law degree distribution.  Networks with
power-law degree distributions occur in many different settings and have
attracted much recent attention~\cite{AB02,DM02,BA99b,ASBS00}.  Percolation
processes on random graphs with power-law degree distributions notably
always have a giant component, no matter how small the percolation
probability~\cite{CEBH00}.  This means for example that a disease will
always spread on such a network, regardless of its transmissibility.  This
result can be modified by more complex network structure such as
correlations between the degrees of adjacent
vertices~\cite{Newman02f,VM03}, but, as we now argue, it is not affected by
clustering.  To see why this is, note that, according to the findings
reported here, we would have to reduce clustering to increase the threshold
above zero, but this is not possible starting from a random graph, which
has $C=0$ to begin with~\cite{WS98}.  ($C$~is fundamentally a probability,
and hence cannot take a negative value.)  Mathematically, we can
demonstrate that our network always percolates using Eq.~\eref{perccond}.
We can create a power-law degree distribution by making the distribution of
number of groups an individual belongs to follow a power law $r_m\sim
m^{-\alpha}$.  (If we wish, we can also make the distribution of group
sizes follow a power law---it doesn't change the qualitative form of our
results.)  The bond occupation probability, and hence the transmissibility,
enters Eq.~\eref{perccond} through the function~$h(z)$, but does not
affect~$f_1(z)$.  We have $f_1'(1)=\sum_m m(m-1) r_m=\av{m^2}-\av{m}$.  For
$\alpha<3$, this diverges, and hence Eq.~\eref{perccond} is always
satisfied, regardless of the value of $p$ or~$T$.

\section{Conclusions}
\label{concs}
We have introduced a solvable model of a network with non-trivial
clustering, and used it to demonstrate, for instance, that increasing the
clustering of a network while keeping the mean degree constant decreases
the size of the giant component.  Increasing the clustering also decreases
the size of an epidemic for an epidemic process on the network, although it
does so at the expense of decreasing the epidemic threshold too.  Among
other things, this means that no amount of clustering will provide us with
a non-zero epidemic threshold in networks with power-law degree
distributions.

\begin{acknowledgments}
The author would like to thank Cris Moore, Juyong Park, Len Sander, and
Duncan Watts for useful and interesting conversations.  This work was
funded in part by the National Science Foundation under grant number
DMS--0234188.
\end{acknowledgments}

\appendix
\section{Probabilities for connected graphs}
Equation~\eref{pkform} implies that we can find a general expression for
$P(k|k)$ if we can calculate the number of connected graphs with a given
number of vertices and edges.  The standard method for counting such graphs
is to write down the exponential generating function for possibly
disconnected graphs and perform an inverse exponential transform to give
the so-called Riddell formula~\cite{RU53}:
\begin{equation}
\sum_{kl} M^k_l {x^k\over k!} y^l
  = \log\biggl(
    1 + \sum_{n=1}^\infty (1+y)^{n(n-1)/2} {x^n\over n!} \biggr).
\end{equation}
Putting $y\to p/q$, $x\to x\sqrt{q}$, and making use of Eq.~\eref{pkform},
we then derive the following generating function for~$P(k|k)$:
\begin{equation}
\sum_{k=1}^\infty q^{-k^2/2} P(k|k) {x^k\over k!}
  = \log \biggl( \sum_{n=0}^\infty q^{-n^2/2} {x^n\over n!} \biggr).
\label{pkgf}
\end{equation}
The sum on the right-hand side is strongly divergent for $|q|<1$, but
progress can be made by allowing $q$ to take a non-physical value greater
than~1 and then analytically continuing to the physical regime.  Using the
fact that the Gaussian is its own Fourier transform:
\begin{equation}
\e^{-t^2/2} = {1\over\sqrt{2\pi}} \int_{-\infty}^\infty
              \e^{-z^2/2} \e^{\i zt} \>\d z,
\end{equation}
the sum can be written~\cite{FSS02}
\begin{eqnarray}
& & \hspace{-4em}
  \sum_{n=1}^\infty {1\over\sqrt{2\pi}} \int_{-\infty}^\infty
  \e^{-z^2/2} \e^{\i zn\sqrt{\log q}} \>\d z\> {x^n\over n!} \nonumber\\
  &=& {1\over\sqrt{2\pi}} \int_{-\infty}^\infty
      \exp\Bigl( -\half z^2 + x\e^{\i z\sqrt{\log q}} \Bigr)\>\d z,
\end{eqnarray}
where we have interchanged the order of sum and integral.

Unfortunately, the integral cannot be carried out in closed form, and
although some asymptotic results can be derived using saddle-point
expansions, it does not appear at present that a closed-form solution for
the generating function $h_n(z)$, Eq.~\eref{defshn}, can be simply derived.


\begin{thebibliography}{10}
\expandafter\ifx\csname url\endcsname\relax
  \def\url#1{\texttt{#1}}\fi
\expandafter\ifx\csname urlprefix\endcsname\relax\def\urlprefix{URL }\fi

\bibitem{Strogatz01}
S.~H. Strogatz, Exploring complex networks. \textit{Nature} \textbf{410},
  268--276 (2001).

\bibitem{AB02}
R.~Albert and A.-L. Barab\'asi, Statistical mechanics of complex networks.
  \textit{Rev. Mod. Phys.} \textbf{74}, 47--97 (2002).

\bibitem{DM02}
S.~N. Dorogovtsev and J.~F.~F. Mendes, Evolution of networks. \textit{Advances
  in Physics} \textbf{51}, 1079--1187 (2002).

\bibitem{AJB99}
R.~Albert, H.~Jeong, and A.-L. Barab\'asi, Diameter of the world-wide web.
  \textit{Nature} \textbf{401}, 130--131 (1999).

\bibitem{FFF99}
M.~Faloutsos, P.~Faloutsos, and C.~Faloutsos, On power-law relationships of the
  internet topology. \textit{Computer Communications Review} \textbf{29},
  251--262 (1999).

\bibitem{ASBS00}
L.~A.~N. Amaral, A.~Scala, M.~Barth\'el\'emy, and H.~E. Stanley, Classes of
  small-world networks. \textit{Proc. Natl. Acad. Sci. USA} \textbf{97},
  11149--11152 (2000).

\bibitem{Newman01a}
M.~E.~J. Newman, The structure of scientific collaboration networks.
  \textit{Proc. Natl. Acad. Sci. USA} \textbf{98}, 404--409 (2001).

\bibitem{Liljeros01}
F.~Liljeros, C.~R. Edling, L.~A.~N. Amaral, H.~E. Stanley, and Y.~\AA{}berg,
  The web of human sexual contacts. \textit{Nature} \textbf{411}, 907--908
  (2001).

\bibitem{WS98}
D.~J. Watts and S.~H. Strogatz, Collective dynamics of `small-world' networks.
  \textit{Nature} \textbf{393}, 440--442 (1998).

\bibitem{CPV03}
G.~Caldarelli, R.~Pastor-Satorras, and A.~Vespignani, Cycles structure and
  local ordering in complex networks. Preprint cond-mat/0212026 (2002).

\bibitem{MR95}
M.~Molloy and B.~Reed, A critical point for random graphs with a given degree
  sequence. \textit{Random Structures and Algorithms} \textbf{6}, 161--179
  (1995).

\bibitem{MR98}
M.~Molloy and B.~Reed, The size of the giant component of a random graph with a
  given degree sequence. \textit{Combinatorics, Probability and Computing}
  \textbf{7}, 295--305 (1998).

\bibitem{ACL00}
W.~Aiello, F.~Chung, and L.~Lu, A random graph model for massive graphs. In
  \textit{Proceedings of the 32nd Annual {ACM} Symposium on Theory of
  Computing}, pp. 171--180, Association of Computing Machinery, New York
  (2000).

\bibitem{NSW01}
M.~E.~J. Newman, S.~H. Strogatz, and D.~J. Watts, Random graphs with arbitrary
  degree distributions and their applications. \textit{Phys. Rev. E}
  \textbf{64}, 026118 (2001).

\bibitem{CL02a}
F.~Chung and L.~Lu, Connected components in random graphs with given degree
  sequences. \textit{Annals of Combinatorics} \textbf{6}, 125--145 (2002).

\bibitem{KE02}
K.~Klemm and V.~M. Eguiluz, Highly clustered scale-free networks. \textit{Phys.
  Rev. E} \textbf{65}, 036123 (2002).

\bibitem{HK02b}
P.~Holme and B.~J. Kim, Growing scale-free networks with tunable clustering.
  \textit{Phys. Rev. E} \textbf{65}, 026107 (2002).

\bibitem{MV03}
Y.~Moreno and A.~V\'azquez, Disease spreading in structured scale-free
  networks. Preprint cond-mat/0210362 (2002).

\bibitem{DGM02a}
S.~N. Dorogovtsev, A.~V. Goltsev, and J.~F.~F. Mendes, Pseudofractal scale-free
  web. \textit{Phys. Rev. E} \textbf{65}, 066122 (2002).

\bibitem{RB03}
E.~Ravasz and A.-L. Barab\'asi, Hierarchical organization in complex networks.
  Preprint cond-mat/0206130 (2002).

\bibitem{GN02}
M.~Girvan and M.~E.~J. Newman, Community structure in social and biological
  networks. \textit{Proc. Natl. Acad. Sci. USA} \textbf{99}, 8271--8276 (2002).

\bibitem{WDN02}
D.~J. Watts, P.~S. Dodds, and M.~E.~J. Newman, Identity and search in social
  networks. \textit{Science} \textbf{296}, 1302--1305 (2002).

\bibitem{Hethcote00}
H.~W. Hethcote, Mathematics of infectious diseases. \textit{SIAM Review}
  \textbf{42}, 599--653 (2000).

\bibitem{Grassberger83}
P.~Grassberger, On the critical behavior of the general epidemic process and
  dynamical percolation. \textit{Math. Biosci.} \textbf{63}, 157--172 (1983).

\bibitem{Sander02}
L.~M. Sander, C.~P. Warren, I.~Sokolov, C.~Simon, and J.~Koopman, Percolation
  on disordered networks as a model for epidemics. \textit{Math. Biosci.}
  \textbf{180}, 293--305 (2002).

\bibitem{Newman02c}
M.~E.~J. Newman, Spread of epidemic disease on networks. \textit{Phys. Rev. E}
  \textbf{66}, 016128 (2002).

\bibitem{BA99b}
A.-L. Barab\'asi and R.~Albert, Emergence of scaling in random networks.
  \textit{Science} \textbf{286}, 509--512 (1999).

\bibitem{CEBH00}
R.~Cohen, K.~Erez, D.~{ben-Avraham}, and S.~Havlin, Resilience of the
  {I}nternet to random breakdowns. \textit{Phys. Rev. Lett.} \textbf{85},
  4626--4628 (2000).

\bibitem{Newman02f}
M.~E.~J. Newman, Assortative mixing in networks. \textit{Phys. Rev. Lett.}
  \textbf{89}, 208701 (2002).

\bibitem{VM03}
A.~V\'azquez and Y.~Moreno, Resilience to damage of graphs with degree
  correlations. \textit{Phys. Rev. E} \textbf{67}, 015101 (2003).

\bibitem{RU53}
R.~J. Riddell, Jr. and G.~E. Uhlenbeck, On the theory of the virial development
  of the equation of state of mono-atomic gases. \textit{J. Chem. Phys.}
  \textbf{21}, 2056--2064 (1953).

\bibitem{FSS02}
P.~Flajolet, B.~Salvy, and G.~Schaeffer, {A}iry phenomena and analytic
  combinatorics of connected graphs. Preprint, {INRIA} (2002).

\end{thebibliography}
\end{document}